\documentclass[prb,preprint]{revtex4-1}

\usepackage[dvipsnames]{xcolor}
\usepackage{amsmath}  
\usepackage{amsfonts} 
\usepackage{graphicx} 
\usepackage{gensymb} 
\usepackage{xcolor}    
\usepackage{epstopdf}
\usepackage{array}
\usepackage{setspace}
\usepackage{amsbsy}
\usepackage{stmaryrd} 
\usepackage{braket}
\usepackage{amsmath}

\begin{document}


\title{Life, death, new species and children from a ``Quantum Game of Life"}

\author{David A. Faux}
\email{d.faux@surrey.ac.uk} 

\author{Mayank Shah}
\author{Christopher Knapp}

\affiliation{Department of Physics, University of Surrey, Guildford, GU2 7XH, UK}


\date{\today}

\begin{abstract}

  The {\it classical} ``game of life" (GOL) due to Conway\cite{gardner_1970,berlekamp1982guy} is a  famous mathematical game constructed as a two-dimensional cellular automaton in which each cell is either alive or dead. A set of evolutionary rules determines whether a cell dies, survives or is born at each generation based on its local environment.  The game of life is interesting because complexity emerges from simple rules and mimics real life in that a cell flourishes only if the environment is ``just right" producing a breadth of life-like behavior including multi-cellular lifeforms.  Results are presented from a quantum adaptation of the GOL which assigns a qubit to each cell which then evolves according to modified evolutionary rules. Computer simulation reveals remarkable evolutionary complexity that is distinct from the classical GOL and which mimics aspects of quantum biological processes and holds promise for the realistic simulation of species population dynamics.  Liveness emerges as a probability density with universal statistical properties dependent solely on the evolutionary rules.  New species of quantum lifeform are found. One quantum lifeform is shown to act as a seed to produce children, one or more classical and/or quantum lifeforms, classical oscillators, a liveness probability density or death with outcomes highly sensitive to the initial state. We observe the emergence of chaos and scaling phenomena making the quantum game of life an exciting prospect for further exploration as a model for life-like behaviors.
\end{abstract}

\maketitle 

\section{Introduction} 

Imagine a two-dimensional universe composed of a square grid of cells. Each cell $i$ in Conway's classical game\cite{gardner_1970,berlekamp1982guy} takes the value $a_i=1$ if alive and $a_i=0$ if dead.  A cell $i$ interacts with the eight cells $j$ in its immediate neighborhood and its evolution depends on the total ``liveness" $A_i=\sum_{j=1}^8 a_j$ of the neighborhood which is an integer from 0 to 8.  A live cell will die at the next time step if $A_i \leq 1$ or $A_i \geq 4$ and survive otherwise.  A dead cell remains dead unless $A_i=3$ in which case a life is born.  These classical rules mimic the success or demise of life based on environment; if the number of live cells in the neighborhood is too high or too low the cell dies due to overcrowding or loneliness but flourishes if the surrounding liveness is optimum.  Conway's evolutionary rules are simple and yet, when applied simultaneously to each cell at each time step called a generation\cite{gardner_1970}, a myriad of complex behaviors emerge.  Multi-cellular lifeforms are created which are classed as ``still lifes" (static or unchanging), ``oscillators" which return to their original state after a number of generations, dynamic objects, plus other more exotic species.  The game also exhibits self-organized criticality with universal scaling laws\cite{bak1989self}.

A process for the quantisation of Conway's classical game was proposed in 2008 by Flitney and Abbott\cite{flitney_abbott_2008} and referred to here as a quantum game of life (QGOL).  In the QGOL, each cell is described by a qubit $\ket{\psi}$ which is a superposition of alive $\ket{1}$ and dead $\ket{0}$ states such that
\begin{equation}
\arraycolsep=1.0pt\def\arraystretch{0.8}
\ket{\psi} = a \ket{1} + b \ket{0}  = \bigg( \begin{array}{c} a \\ b  \end{array} \bigg)
\end{equation}
where $a$ is the ``liveness" of the cell and $b$ is obtained by normalisation. If a cell $i$ is tested to establish its liveness, it will be found to be alive with a probability $|a_i|^2$ or dead with a probability $|b_i|^2$. Flitney and Abbott \cite{flitney_abbott_2008} proposed adapted evolutionary rules that modify the liveness of a each cell $i$ at each generation dependent on its coupling to the neighborhood liveness  $A_i$, now a continuous variable between 0 and 8.  The QGOL rules recover Conway's classical game for purely live and dead cells. The QGOL rules are non-unitary (application of the QGOL rules does not conserve liveness probability density) and so each application of the QGOL rules includes a normalisation step.  
Each cell $\ket{\psi_i}$ is therefore a qubit which interacts with its immediate environment $A_i$ at each generation to evolve to a new state $\ket{\psi_i^\prime}$.  Qubits do not exhibit quantum coherence nor execute random walks but increase, decrease or maintain their liveness solely in response to the local environment.  The QGOL therefore has a parallel in quantum biology whereby an exciton is produced by a photon in a protein complex during photosynthesis\cite{engel2007evidence}. The exciton interacts with its environment to influence the energy transfer dynamics\cite{mohseni2008environment}. The quantum cellular automaton studied here constitutes a simple cellular model of biological systems containing a quantum objects whose states are modified in response to their immediate surroundings.

\begin{figure}	[tbh!]		
	\includegraphics[width=15. cm, height=11.0cm, trim={2cm 1.5cm 3cm 1cm},clip]{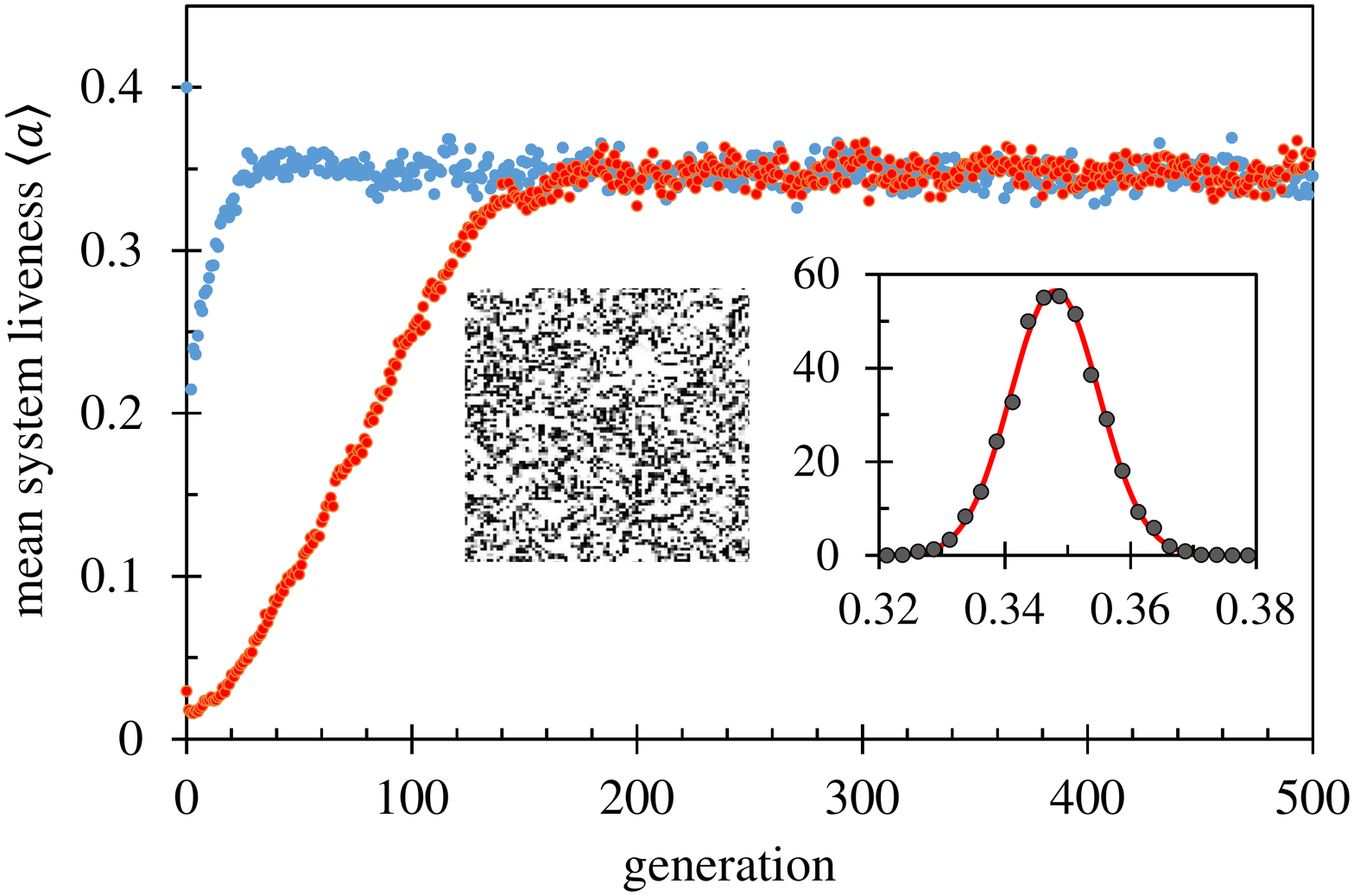}
	\caption{The mean cell liveness $< \! a \! >$ of a QGOL system of $100 \times 100$ cells with a starting fraction of live cells $f=0.2$ (\textcolor{red}{\pmb{$\bullet$}}) and $f=0.8$ (\textcolor{RoyalBlue}{\pmb{$\bullet$}}) is presented. The inset shows the Gaussian distribution of $< \! a \! >$. An example cell liveness distribution is presented as a gray-scale (dead=white, live=black).}
	\label{Fig1_eqm}
\end{figure}

The QGOL was executed on a $100 \times 100$ grid with periodic boundary conditions.  To initiate the system, a fraction $f$ of cells was assigned a liveness chosen via a random number uniformly distributed between 0 and 1. The value of $b$ for each cell is obtained by normalisation.  The remaining cells are assigned as dead, that is $a=0, b=1$.  In this work we restrict the qubit to real components. Flitney and Abbott\cite{flitney_abbott_2008} briefly examined a one-dimensional system of oscillating qubits but the simpler model presented here is previously unexplored.

The system is evolved according to the QGOL rules of Flitney and Abbott\cite{flitney_abbott_2008} to reach a dynamic equilibrium comprising dead cells, live cells and cells in superposed, or part-live, states as shown in Fig.~\ref{Fig1_eqm}.
The mean cell liveness is 
$< \! a \! >=0.3480 \pm 0.0001$ and the distribution of mean liveness is Gaussian with a standard deviation 0.0071 as shown in Fig.~\ref{Fig1_eqm}. The mean and standard deviation are ``universal constants" in the sense that they are independent of starting conditions and dependent solely on the underlying evolutionary rules.

\begin{figure}	[tbh!]	
	\includegraphics[width=11cm, height=12.0cm, trim={0cm 18cm 12cm 1cm},clip]{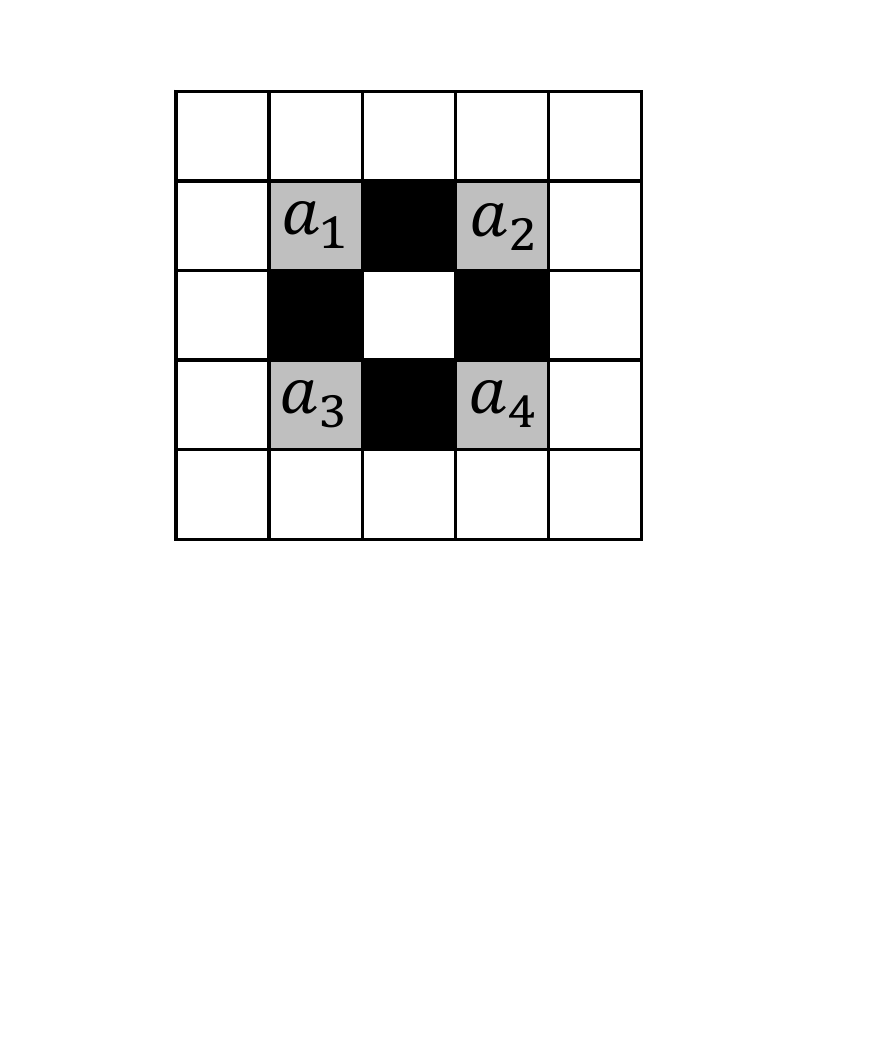} \\[-5cm]
	\caption{The qutub structure is presented. The qutub is unchanging if nearest-neighbor part-live cells sum to one or less ($a_1 + a_2 \leq 1$ etc.)}
	\label{Fig2_qutub}
\end{figure}

An examination of dynamic liveness distributions reveals transient multi-cellular lifeforms emerging temporarily from the liveness cloud. These lifeforms exist for a small number of generations before encroachment and destruction by nearby density. The lifeforms may be classical lifeforms seen in Conway's GOL or previously-unseen multicellular quantum lifeforms.  The quantum lifeforms comprise a combination of live ($a=1$) and/or dead ($a=0$) cells and cells in part-live states represented by $0 < a < 1$.  The quantum lifeforms are static (unchanging) when placed in an otherwise empty universe provided certain criteria based on the values of $a$ are satisfied.  We have identified 12 species of multi-cellular quantum lifeform.  The ``qutub" (named empathetic to its classical counterpart) is of special interest and is illustrated in Fig.~\ref{Fig2_qutub}.  The qutub is unchanging if the livenesses $a_i$ of any pair of next-nearest (non-diagonal) neighbors sum to 1 or less.  The qutub is dormant, or hibernating, until an event occurs such that the stability criteria is no longer satisfied in which case the qutub acts as a seed to spawn a variety of structures depending on the changed livenesses $a_i$ of the four part-live cells.  

\begin{figure}	[tbh!]	
	\centering
	\includegraphics[width=10. cm, height=16.0cm, trim={0cm 2cm 5cm 1cm},clip]{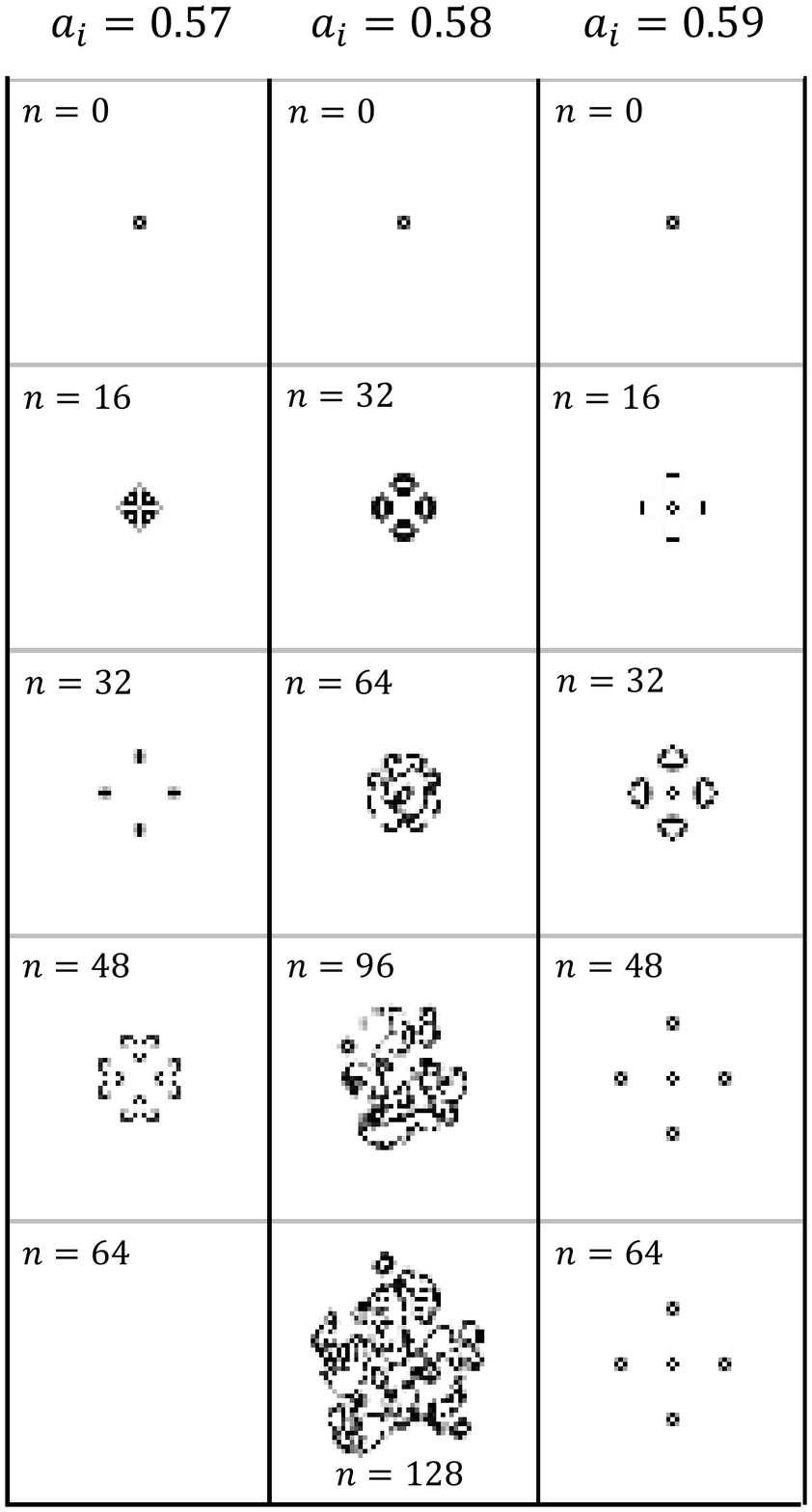}
	\caption{The evolution of a single qutub with $a_1=a_2=a_3=a_4$ equal to 0.57, 0.58 or 0.59 are presented. Snapshots at generation $n$ are shown.} 
	\label{Fig3_evolutions}
\end{figure}

A qutub was placed at the center of an otherwise dead universe and used as a seed to spawn new evolutionary systems. 
The variety of outcome and sensitivity to the initial conditions is illustrated in Fig.~\ref{Fig3_evolutions} for three qutub seeds in which $a_{1...4}=$0.57, 0.58 or 0.59.   If $a_{1...4}=0.57$, the system ultimately dies.  For $a_{1...4} = 0.58$, the qutub seed evolves to produce a space-filling liveness density.  This liveness cloud possesses the same mean and standard deviation as seen in Fig~\ref{Fig1_eqm}. For the case $a_{1...4} = 0.59$, the evolution produces a central classical tub structure plus four child qutubs. 

The qutub seeds in Fig.~\ref{Fig3_evolutions} are initiated with four-fold liveness symmetry since each $a_i$ takes the same starting value.  Four-fold evolutionary symmetry would therefore be expected. This does not arise. This is clear in the evolution of liveness density in Fig.~\ref{Fig3_evolutions} and the quadruplets are also non-identical (akin to human identical twins which have identical DNA but nonetheless possess individual distinguishable features). The loss of four-fold symmetry is due to the inability to present, with precision, a floating point number as a finite-length bit string. The computations use single-precision floating-point computation on a workstation and the computations at each evolutionary step differ slightly due to the order in which the neighborhood sums are executed.  The small, unpredictable differences in floating-point calculations in the quantum game propagate through the generations and become exaggerated due to the evolutionary rules.  The consequence is that the evolution develops ``chaotically" as if the imprecise seventh or eighth decimal digit acts as the butterfly wings.  This feature is not seen in the classical game which uses integer arithmetic.  

\begin{figure}	[tbh!]	
	\includegraphics[width=15. cm, height=14.0cm, trim={0cm 10cm 0cm 1cm},clip]{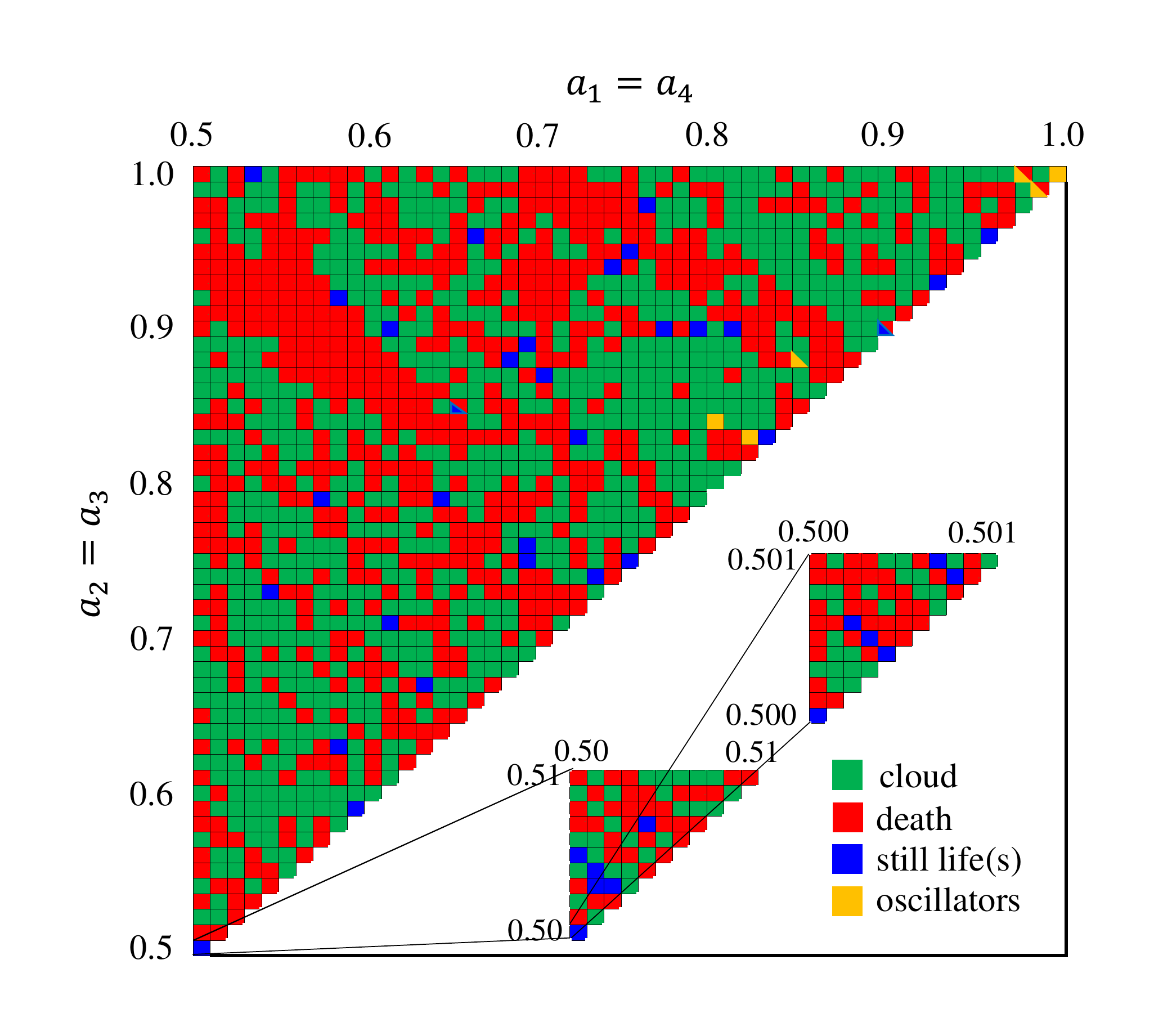}
	\caption{The evolutionary outcomes of a qutub seed with initial livenesses $a_i$ are presented. Half-cells indicate temporary forms. }
	\label{Fig4_colorgrid}
\end{figure}

The evolution of qutub seeds with two-fold symmetry is explored by setting diagonally-opposite part-live cell to the same value such that $a_1=a_4=0.5 ... 1.0$ and $a_2=a_3=0.5 ... 1.0$ at intervals of 0.01. The leading diagonal represents $a_1=a_2=a_3=a_4$.  The evolutionary outcomes presented in Fig.~\ref{Fig4_colorgrid} summarise the complexity of outcome and the sensitivity to the initial state. Systems may die, or produce an expanding liveness cloud. Some qutub seeds evolve to form child qutubs, other quantum lifeforms or combinations.  One case leads to 16 classical oscillators, others to the classical tub structure or combinations of classical and quantum lifeforms.  Occasionally lifeforms exist for more than 4 generations before eventually evolving to life-cloud or death indicated as half squares in Fig.~\ref{Fig4_colorgrid}.  

The fine detail of the qutub evolution reveals a further interesting feature.  If $a_1=a_2=a_3=a_4=0.5$, the qutub is a still-life.  This is represented by the blue square at bottom left in Fig.~\ref{Fig4_colorgrid}. An exploration of the part-liveness parameter space at finer divisions of 0.001 for the range 0.50..0.51 and divisions of 0.0001 for the range 0.500..0.501 are shown in Fig.~\ref{Fig4_colorgrid}.  These yield a broadly similar distribution of life-cloud, death and still-life outcomes. Qualitatively, the system exhibits scaling behavior with similar properties at different scales and additionally serves to emphasise the fragility of life through the extreme sensitivity to starting conditions.

Games of life are not designed to simulate specific biological systems. They simulate evolutionary processes by the application of simple environment-dependent local rules  to produce multi-cellular lifeforms and complex behaviors.  The QGOL is vastly different to its classical counterpart with life described by probability density and arguably provides a more physically-relevant simple game-based model for many real-life processes than its classical counterpart.  Distributions mimic the evolution of moulds, bacterial populations and may with adaptation characterise the fluctuating fortunes of species.  The QGOL provides an expanse of simulation opportunities in unexplored game territory.

\bibliography{References_QGOL}

\end{document}